\documentclass[11pt]{article}

\usepackage{epsfig, amsfonts, amsmath, amssymb, amsthm, pictex, verbatim, amsthm, pdfsync, mathrsfs}
\usepackage{tensind}
\usepackage{endnotes}
\usepackage[normalem]{ulem}

\usepackage{hyperref}  %This package should often be loaded last, to prevent conflict with other packages.
\hypersetup{colorlinks=true, citecolor=blue, urlcolor=blue,linkcolor=blue} 
\usepackage{graphicx}
\usepackage{amsmath}
\usepackage{cases}
\usepackage{amssymb}
\usepackage{hyphenat}
\usepackage[hypcap]{caption} %When hyperrefs link to figures, they go to the top of the figure, instead of to the caption.

\usepackage{tikz} 					%Drawing pictures within LaTeX.
\usetikzlibrary[patterns]				% Different patterns to fill areas.
\usetikzlibrary[arrows]				% Different arrow heads
\usetikzlibrary{positioning}			% Rotating (e.g. vertical axis labels)
\usepackage[color=yellow, textwidth=2.8cm, textsize=tiny, shadow, disable]{todonotes}				%Inserting todo notes. One option that I usually use: colorinlistoftodos %Change colour of bublles and remove lines: \todo[noline,color=red]{comment}
\usepackage{marginnote}

\usepackage{framed}			%The package creates three environments:framed, which puts an ordinary frame box around the region,shaded, which shades the region, and left­bar, which places a line at the left side. The environments allow a break at their start (the \FrameCommand enables creation of a title that is “attached” to the environment); breaks are also allowed in the course of the framed/shaded matter.

\usepackage{comment}

\usepackage{booktabs} % for much better looking tables
\usepackage{multirow}
\usepackage{tabularx}
\usepackage{fixltx2e} % Allows for \textsubscript
\usepackage{booktabs} % for much better looking tables
\usepackage{multirow}

\usepackage{verbatim}  % Required for the automatic word count.

\usepackage{caption}
\usepackage{subcaption}   % These two packes (caption and subcaption) are used to get nice arrays of figures.

\usepackage{cite}  % [1,2,3]  will become [1-3].

\usepackage[margin=1.52in]{geometry}
\usepackage{authblk}

\newcommand*\makeAlph[1]{\symbol{\numexpr64+#1}}
\usepackage{xparse}

\makeatletter
\newcommand{\customlabel}[2]{ \protected@write \@auxout {}{\string \newlabel {#1}{{#2}{\thepage}{#2}{#1}{}} } \hypertarget{#1}{}}
\makeatother
\definecolor{cblue}{RGB}{100,5,255}    
\definecolor{cred}{RGB}{255,10,10} 
\definecolor{cgreen}{RGB}{5,165,20}  
\definecolor{corange}{rgb}{1.0,0.49,0.0}  
                                                                   
\newcommand{\niels}[1]{{\color{cred}#1}}    

\newcommand{\cniels}[1]{\emph{\color{cred}[#1]}} 

\newcommand{\g}{g_{\mu \nu}}

\newcommand{\T}{T_{\mu \nu}}

\title{Dark Matter = Modified Gravity? \\ \Large Scrutinising the spacetime--matter distinction through the modified gravity/ dark matter lens}
\author[1,2,3]{Niels C.M.\ Martens}
\author[1,2]{Dennis Lehmkuhl}
\affil[1]{DFG Research Unit ``The Epistemology of the Large	Hadron Collider'' (grant FOR 2063)\newline \url{martens@physik.rwth-aachen.de}, \url{dennis.lehmkuhl@uni-bonn.de}}
\affil[2]{Lichtenberg Group for History and Philosophy of Physics, University of Bonn}
\affil[3]{Institute for Theoretical Particle Physics and Cosmology, RWTH Aachen University\vspace{3mm} \newline \normalfont{This is a postprint (i.e.~post-peer-review but pre-copyedit) version of an article accepted for publication in} \textit{Studies in History and Philosophy of Modern Physics}. \normalfont{The final authenticated version is/will be available online (open access) at: \url{www.sciencedirect.com/science/article/abs/pii/S135521982030109X}.}}
\begin{document}
	
\maketitle

\abstract{This paper scrutinises the tenability of a strict conceptual distinction between space(time) and matter via the lens of the debate between modified gravity and dark matter. In particular, we consider Berezhiani and Khoury's novel `superfluid dark matter theory' (SFDM) as a case study. Two families of criteria for being matter and being spacetime, respectively, are extracted from the literature. Evaluation of the new scalar field postulated by SFDM according to these criteria reveals that it is as much (dark) matter as anything could possibly be, but also---below the critical temperature for superfluidity---as much (of a modification of) spacetime as anything could possibly be. A sequel paper examines possible interpretations of SFDM in light of this result, as well as the consequences for our understanding of (the importance of) the modified gravity/ dark matter distinction and the broader spacetime--matter distinction.}

\newpage
\tableofcontents

\vspace{8mm}

\section{Introduction}

If one goes to almost any talk directed at a general audience concerned with cosmology and astrophysics, the speaker is likely to declare that one of the biggest mysteries of contemporary physics is that only 4 or 5\% of the matter in the universe is `normal' matter, the luminous matter we see all around us, 21-25\% is dark matter and the rest is dark energy. Let's bracket the question of whether dark energy really belongs into that list, and whether it really is energy in the sense in which both normal and dark matter posses energy. What, one might ask, has led to the conviction that there is roughly 5 times as much dark matter in the universe as the kind of matter we see all around us, especially given that we have neither directly detected nor produced a dark matter particle on the Earth? Why do we believe in the existence of dark matter?

Well, the truth is that not everybody does. There is a much smaller community of physicists that see themselves in direct opposition to the dark matter cartel. They say that it would be too hasty to conclude that dark matter exists, that indeed the observational data that has led the majority of physicists to this conviction could be accounted for equally well, if not better, by sticking to the belief that the majority of matter in the universe is luminous baryonic matter, and by modifying instead the laws of gravity and spacetime physics. True, members of this camp are less often invited to give public lectures to a general audience. But they have produced impressive results, accounting for the rotation curves of galaxies without the introduction of a new kind of matter. Of course, this does not mean that they are right, either.

The fact of the matter is that the conjunction of the assumptions i) General Relativity (GR), and Newtonian Gravity as its non-relativistic limit, is the correct theory of gravity (and spacetime) and ii) most\footnote{Except for, e.g.~Standard Model neutrinos, but we know that their masses and cosmic densities are not sufficient to account for what we below call the `dark phenomena'.} of the matter in the universe is luminous baryonic matter (the stuff that stars and planets consist of), leads to predictions that have been falsified by observations \cite{sanders2010,bertone2018}. We will call these observations `dark phenomena' or `dark discrepancies'.\footnote{Many in the literature speak of the `missing mass problem', but we chose not to follow this terminology because it seems to presuppose (rather than to argue) that premise ii) above is wrong.} % XXX add a reference to a review that uses `missing mass problem' to this footnote. 
These discrepancies show up at a large range of scales. Starting with the smallest scale: Galaxy rotation curves---orbital speeds of visible stars or gas vs.\ radial distance from the galactic centre---are expected to decline towards the edge of the galaxy, but remain constant. Whole galaxies move faster than is to be expected from the gravitational pull of the galaxy cluster that they are part of. Gravitational lensing of either galaxies or galaxy clusters indicates a stronger bending of light than expected from the mass of the luminous matter. Finally, problems appear even at the cosmological scale. Density fluctuations of luminous matter in the early universe are washed out as long as that matter is in thermal equilibrium with photons. Only after freeze-out can density fluctuations grow under the influence of gravity, suggesting that structure formation should not have progressed as much as we do in fact observe it to have progressed at current times.

Essentially, the difference between the Dark Matter (DM) community and the Modified Gravity\footnote{The common terminology `modified gravity' is unfortunate in our context, in that one of the things at stake here is exactly whether the gravitational field (in relativistic theories) is to be classified as part of the matter or part of the spacetime sector---if such a classification is possible in the first place. Nevertheless, we will in the first instance adopt the standard terminology, rather than, for instance, `modified spacetime'.} (MG) community is that, in response to these discrepancies, the former gives up premise ii) and introduces dark matter as the main kind of matter in the universe, whereas the latter gives up on premise i) and introduces different gravitational field equations.

On galactic scales, using GR over its Newtonian limit likely makes no observable difference for the predictions concerning rotation curves. Thus, the first proposal of the MG community consisted of a modification of the Newtonian gravitational equations, in the form of Mordehai Milgrom's Modified Newtonian Dynamics (MOND). One way of presenting the theory is as a modification of Newton's inverse square law of gravity, which is replaced by:
\begin{numcases}{F_G = G \frac{Mm}{\mu(\frac{a}{a_0}) r^2}
	\label{E:Milgrom gravity} \quad }
\mu \approx 1, & if $a \gg a_0$\\
\mu \approx a/a_0, & if $a \ll a_0$
\end{numcases}
where $a_0$ is a new constant of nature, which has been empirically determined to be $1.2 \times 10^{-10} \frac{m}{s^2}$ \cite{li2018,mcgaugh2018}. This law approximates Newton's law for gravitational accelerations much larger than $a_0$. However, at accelerations much smaller than $a_0$ the gravitational force is enhanced in comparison to Newton's law. It follows that there is no practical difference for the planets in our solar system, but there is for stars in the outer regions of their galaxy. Once this value for $a_0$ is obtained, MOND uniquely and adequately predicts the observed galaxy rotation curves.  

As noted above, MOND described thus seems a clear example of an MG theory, which modifies premise i) while leaving premise ii) untouched. 

Likewise, paradigmatic DM theories can easily be introduced within the Newtonian regime. Choosing (by hand) a suitable distribution of additional, electrically neutral, massive Newtonian objects obeying the (unmodified) laws of Newtonian Gravity can also account for the observed rotation curves of individual galaxies.

At the very least at the cosmological scale, and in the context of gravitational lensing also at the level of galaxies and galaxy clusters, we have to move beyond (modified or unmodified) Newtonian physics and use a relativistic theory---and sometimes also a quantum theory---that is approximated by (modified or unmodified) Newtonian gravity at low velocities and for weak gravitational fields \cite{famaey2012}. The question that we want to address in this paper is whether a clear distinction remains between modifying the gravitational interaction/ spacetime structure (premise i) on the one hand and modifying the assumption of what the matter content in our universe consists of (premise ii). In other words: once we enter the relativistic (and quantum) regime, is there still a clear distinction between gravitational fields and matter fields?

This distinction between gravity/spacetime and matter has been questioned already in the context of classical GR,\footnote{Skepticism about the space(time)--matter distinction may be traced back even further. Around the very end of the 19th century, it was widely held that ``the question whether the aether is really a special form of matter or really just space endowed with certain physical properties is not one of serious consequence, but only a question of what mode of expression one prefers to adopt'' \cite[p.290]{rynasiewicz1996}. Of course the concept of the aether, at least in any of its historical forms, has long been abandoned, but one might wonder---as Rynasiewicz does, and we will do, in the main text---to what extent the (manifold plus) metric field in GR is not a new aether in disguise, in the sense of having both properties that are historically associated with matter and with spacetime.} nevermind dark phenomena. For instance,\footnote{See also \cite{feynman1995,rovelli1997,dorato2000,slowik2005,dorato2008,lehmkuhl2008,rovelli2010,rey2013} \cite[p.354]{vassallo2016} \cite[p.36]{lehmkuhl2018} \cite[\S 3.3]{belot2011b} \cite[\S 8]{greaves2011}.} Earman and Norton argued that the metric field in GR, the successor of the Newtonian gravitational potential in the context of GR, should be seen as belonging to the same ontological category as what represents matter in the context of GR: both are fields defined on the spacetime manifold $M$ \cite{earman1987}.\footnote{See also \cite{anderson1967,rovelli1997} \cite[Ch.9]{brown2005}.} Earman and Norton took inspiration directly from Einstein, who did not see the metric field as categorically different from the electromagnetic field, or indeed any other fields, and staunchly argued that the claim that GR had `geometrized' the gravitational field (but not yet the electromagnetic field) did not have any meaning at all \cite{lehmkuhl2014}. Similarly, Rynasiewicz argued that the old debate between substantivalism and relationalism (the question of whether space or spacetime is a fundamental entity in its own right or derivative of the relationships between material bodies) is ill-founded in the context of GR, primarily because it is just not clear that the metric field $\g$ should be seen as corresponding to (aspects of) spacetime (structure) \cite{rynasiewicz1996}. In a similar vein, one of us has argued that \emph{the} property that makes a field a \emph{matter} field in the context of relativistic field theories is that it has a mass-energy-momentum tensor $\T$ asssociated with it, but that the definition of this matter-making property definitionally depends on $\g$ in a variety of ways  \cite{lehmkuhl2011}. This, too, suggests that the line between `matter' and `gravity / spacetime structure' is not as sharp as has often been claimed.\footnote{We will return to the question of whether `possessing' mass-energy-momentum  is the right criterion for defining `matter' in section \ref{energeticcrit}.} 

Finally, of course, the distinction also comes under fire in various approaches to solving the problem of quantum gravity.\footnote{A particularly interesting context would be string theory: is the dilaton best categorised as spacetime, matter, both or neither? We would like to thank an anonymous reviewer for this suggestion.} However, in this paper we are interested in the degree to which this distinction becomes problematic even before reaching the regime were such theories of quantum gravity\footnote{Although the case study in this paper, superfluid dark matter theory, is a theory of quantum gravity in the sense that a (macroscopic) quantum effect, namely a superfluid Bose-Einstein condensate, leads to a modification of gravity (at the scale of galaxies), it is not a theory of quantum gravity in the sense of solving the problem of quantum gravity (which concerns a regime that is not standardly associated with the galactic scale).} are expected to reign---this has the additional benefit of staying closer to experiment/observation.

In these two companion papers, we shall provide a new argument against the idea that in relativistic and/or quantum theories the distinction between matter and spacetime, and related to that the distinction between modifying the matter content or modifying the gravitational laws, is clear-cut. We shall question the tenability of the spacetime/matter distinction by scrutinising it through the lens of the modified gravity/ dark matter distinction (MG/DM)\footnote{Other interesting pre-quantum-gravity contexts in which one can question the tenability of the spacetime/matter distinction---thereby further contributing to the cartography research programme (see main text and \cite{martenslehmkuhl2})---include spin-2 gravity, Newton-Cartan theory, f(R) gravity and Jordan-Brans-Dicke theories \cite{duerrms}, the cosmological constant, black holes, unified field theories and supersymmetric theories.}---which is, in some senses, more clear, more straightforward than the GR lens. %\todo{Add some of the ``cartography'' stuff, and all the other stuff in the new section 7 in paper 2.}
This is one of many possible ways in which to make a small contribution to the `cartography research programme'. This research programme is concerned with navigating the space of theories via a dynamic back-and-forth between, on the one hand, understanding and interpreting individual theories and, importantly, the relations between neighbouring theories and the concepts they use, and, on the other hand, exploring the larger space of theories (with the help of the lessons learnt from the individual case studies) in order to understand that space as a whole and to streamline theory generation.\footnote{These companion papers thereby cohere with an argument by one of us that (even) if one (only) wants to understand and interpret a (single) theory, in particular our `best' spacetime theory, GR, and figure out what makes it special, one (still) needs to look at the neighbourhood of that theory in the space of spacetime theories.  See \cite{lehmkuhl2017} for details, and for a strategy in a similar spirit see \cite{fletcher2018}.} (More on this in the sequel paper \cite{martenslehmkuhl2}.)

Deflating the strict dichotomy or even the distinction between dark matter theories and modified gravity theories might contribute to undermining the mutual hostility that is prevalent between the two camps that have formed in response to one of modern physics' most pressing problems: the dark discrepancies. Deflating the broader distinction between spacetime and matter could, quite literally, make the oldest foundations of natural philosophy crumble, including the debate that has dominated the philosophy of space(time) over the past three centuries: the substantivalism/relationalism debate.\footnote{A third option in addition to substantivalism and relationalism that is often swept under the carpet is that of super-substantivalism. Where relationalism argues that space(time) is derivable from matter, super-substantivalism argues that matter is derivable from space(time). However, most versions of super-substantivalism do not thereby identify the two notions, ontologically, let alone conceptually. Rynasiewicz argues that not even the arch-super-substantivalist Descartes' definintion of matter as \emph{res extensa} was intended to destroy the conceptual distinction \cite[p.281-2]{rynasiewicz1996}. Skow \cite{skow2005} and Lehmkuhl \cite{lehmkuhl2018} distinguish modest and radical supersubstantivalism. The former is the metaphysical view that, regardless of the specific physical theory (albeit most naturally in a field theory \cite[p.298-9]{rynasiewicz1996}), one can simply stipulate that paradigmatic matter properties such as mass and colour are all directly attributed to spacetime. The latter is the physics research programme of determining to what extent specific physical theories, such as Kaluza-Klein theory or Wheeler's geometrodynamics, manage to reduce these apparent non-geometrical properties to (or have them emerge from) geometrical (or topological) properties. In a sense this paper concerns something even more radical: the research programme of determining to what extent the neighborhood of our best theories in the space of theories suggests a \emph{conceptual} identification, or at least a blurring, of the orthodox distinction between spacetime and matter.}

We will go about this by investigating a particularly promising theory due to Lasha Berezhiani and Justin Khoury \cite{berezhiani2015,berezhiani2016}.\footnote{For potential further interesting DM/MG case studies, see \cite{bekenstein2004,blanchet2008,zhao2008,bruneton2009,li2009,ho2010,ho2011,ho2012,cadoni2018,cadoni2019,scholz2020,ferreira2020,skordis2020}.} They call it ``superfluid dark matter theory''; we shall call it SFDM in the following. Their theory comes as a response to the stalemate between the DM and the MG research programmes---DM theories have traditionally done well at the level of cosmology and galaxy clusters, but less so at the level of galaxies,\footnote{This may seem to contradict our earlier claim that one can account for the observed rotation curves of individual galaxies by a suitable distribution of dark Newtonian objects obeying the (unmodified) laws of Newtonian gravity. It is indeed true that if one models the dark matter distribution with sufficiently many parameters and adjusts them by hand for each individual galaxy, that one can fit the observed rotation curve of that galaxy. However, if one does not put in each galactic distribution by hand, but uses the output of computer simulations of the evolution of the whole universe, one has trouble reproducing the correct locations, numbers and shapes of dark matter halos, as well as several empirical correlations across galaxies, such as the baryonic Tully-Fisher relation \cite{delpopolo2017}.} with the opposite being the case for MG theories. Superfluid dark matter theory has gained quite some attention in the physics community, particularly because it seems  that given SFDM you can have your cake and eat it: in their seminal paper of 2015, the authors announce that SFDM ``matches the successes of the $\Lambda$ cold dark matter ($\Lambda$CDM) model on cosmological scales while simultaneously reproducing the modified Newtonian dynamics (MOND) phenomenology on galactic scales'' \cite[p.1]{berezhiani2016}. The way in which this is achieved is by introduction of a new self-interacting, massive and complex scalar field $\Phi$ which the authors classify as a dark matter field. However, $\Phi$ has two phases corresponding to the two domains in which typical DM and MG theories are successful; galaxy clusters and galaxies, respectively. On the length scale of galaxy clusters, $\Phi$  is in the normal fluid phase, giving rise to axion-like particles with masses in the range of eVs and strong self-interactions. This allows for reproduction of the successes of typical dark matter theories. On the length scale of single galaxies, $\Phi$ enters a superfluid phase, giving rise to superfluid phonons that are governed by a MOND-like action (in the non-relativistic regime). It is thus that they can reproduce MOND phenomenology in galaxies. Borrowing terminology from an approach of kindred spirit due to Sabine Hossenfelder \cite{hossenfelder2017}, one might call $\Phi$ in the superfluid phase an ``imposter field'': a field that is `really' dark matter, but acts \emph{as if} it was a modification of gravity, a new gravitational degree of freedom, in a certain domain. (Interestingly, Hossenfelder refers to SFDM as a variant of modified gravity. SFDM is formally very similar to her own covariant version of Verlinde's emergent/entropic gravity. She interprets the new field in her own theory as a modification of gravity, with plays the role of dark matter without `being' dark matter. It is hence considered an imposter field (but in the sense opposite to that above).)

The aim of this pair of companion papers will be to unpack and analyse such interpretational claims. Is $\Phi$ `really' a dark matter field and only acts like a modification of gravity in certain regimes? Or should one say that it is fundamentally a new gravitational field that acts like a dark matter field in certain regimes? If not, why not? Or is the sense in which $\Phi$ is dark matter on a par with the sense in which it is a modification of gravity and spacetime? And indeed, what are the criteria that justify calling a scalar field (and its excitations) a dark matter field in the first place, or a new gravitational field / gravitational degree of freedom? Or does SFDM put pressure on the contemporary usefulness or even the coherence of these old categories?

We will proceed in the following way. In this first paper, we will discuss the criteria that might be put forward to categorise a newly introduced tensor (or spinor) field as a matter field (section \ref{S:kinetic}) or an aspect of spacetime (section \ref{S:geometric}), respectively. In the latter case, we will focus on the question of what counts as modifying the representation of gravitational fields as compared to GR, which is typically the starting point of Modified Gravity approaches. Since in GR the representation of gravity is based on pseudo-Riemannian geometry, many modifications of GR are modifications of that underlying geometrical structure. However, note that we do not thereby commit to the claim that every theory of gravity will conceive of gravity as something necessarily connected to spacetime geometry; indeed, this opinion is controversial even in the context of GR. In short, some but not all theories of gravity are spacetime theories, just like some but not all spacetime theories are theories of gravity. Still, within the realm of modified gravity theories aiming to give an account of the dark discrepancies, the decisive change as compared to GR is typically a geometrical one; as we shall see, SFDM, by following a prominent relativistic extension of MOND called TeVeS, involves the introduction of a second, `physical' metric tensor, and a coming apart of the different roles the metric tensor plays in GR.%\todo[color=white]{\niels{Perhaps we should explicitly mention these different roles somewhere in the paper. What do you take those roles to be?}}

Following the discussion of the criteria that might lead one to consider a new field a matter field or a modification of spacetime and gravity (or a new gravitational and/or spatiotemporal degree of freedom), we shall---after introducing SFDM in more detail (section \ref{FDMintro})---apply these criteria to the case of SFDM (section \ref{S:FDMeval}). We shall argue that the often unquestioned assumption that every field is either a matter field or a gravitational/spacetime field is put under severe pressure by SFDM. In particular, we will argue that the newly introduced scalar field in SFDM is both a matter field and a gravitational/spacetime field. This result will then be the starting point of the second companion paper \cite{martenslehmkuhl2}, where we will unpack the interpretational questions raised above, both in the context of SFDM as well as in the broader context of charting (the neighbourhood of SFDM in) the space of theories.

%The  Though the authors introduce the theory as a dark matter theory, we shall argue that the theory has just as many properties typically associated with a modified gravity theory. What is more, there are equally strong arguments for seeing the new field introduced by the theory as a dark matter field as there are for seeing it as a new gravitational degree of freedom. As we shall see, this leaves us with three options: either we say that the theory consists neither in  an introduction of dark matter nor in a modification of the gravitational degrees of freedom; or we say that it consists in both.

\section[Matter Criteria]{Matter Criteria} \label{S:kinetic}

The first dominant family of criteria in the literature are often considered the litmus test for calling something matter.\footnote{\label{physicalvsmatter}Sometimes some of these criteria are used as criteria for being physical---for being real \cite[Chs.5,8]{lange2002} \cite[\S 4.2]{lazarovici2018} \cite{bunge2000,sebensforth,frisch2005}---rather than (just) being matter. However, equating `matter' and `physicality'/`reality' would imply that spacetime must either be matter or it must be unphysical. We take it that there is nothing unphysical about pure spacetime (i.e.\ something that is spacetime but not simultaneously matter), such as Newtonian spacetime. We shouldn't \emph{a priori} rule out the possibility of our world (or other worlds) containing pure, physical spacetime.}$^,$\footnote{Baker takes the negation of some of these criteria as counting toward an object satisfying the spacetime concept \cite{bakerstfunc}.} Explicit discussions of a (naturalistic) definition of matter (as opposed to spacetime) are surprisingly rare. Interestingly, one of the clearest discussions of these criteria occurred when Earman and Norton's hole argument \cite{earman1987} lead to a discussion on the concept of \emph{spacetime}. The ensuing debate targeted a conception of substantival spacetime that excluded the metric field:  

\begin{quote} \label{earmannortonquote}
Norton and Earman provide several considerations designed to promote the bare manifold as the most plausible candidate for substantival space-time. The leading arguments are: a) In the GTR [General Theory of Relativity] the metric field is just like any other field, both mathematically and physically. It is represented by a tensor similar to, e.g., the electromagnetic field; it is governed by local differential equations; it carries energy and momentum. Hence it should not be essential to space-time (Earman \& Norton 1987, p. 519). b) In the GTR the metric becomes a dynamical object. Hence it should not be considered essential to space-time (Earman forthcoming, chapter 10). \cite[p.97]{maudlin1988}
\end{quote}
This line of reasoning could be interpreted as the metric field being just like any other \emph{matter} field---which would make the mentioned properties aspects of the definition of matter.\footnote{Another possible interpretation however, closer to Einstein (and arguably intended by Earman and Norton), is that the metric field and paradigmatic `matter' fields such as the electromagnetic fields are all just fields, with the further distinction between matter fields and other fields being  unimportant, or even misleading. Einstein's position on this will be spelt out and discussed in detail in \cite{Lehmkuhl:forthOUP}.} Rovelli uses the term `matter' explicitly:
\begin{quote}
	[I]n the general relativistic world picture [the spacetime-versus-matter distinction] collapses. In general relativity, the metric/gravitational field has acquired most, if not all, the attributes that have characterized matter (as opposed to spacetime) from Descartes to Feynman: it satisfies differential equations, it carries energy and momentum, and, in Leibnizian terms, it \emph{can act and also be acted upon}, and so on. \cite[p.193]{rovelli1997}\footnote{See also \cite{rovelli2010}.}
\end{quote}

Let us pick these matter criteria, often put forward as necessary and/or sufficient criteria for being matter, apart systematically, and order them according to logical strength. Initially, their differences will be illustrated by applying them, as in the above quotes, to the metric field in GR. In some cases this application is controversial. We will see below that in the context of MG and DM theories their application is more straightforward.

\subsection{Kinetic (matter) criteria}

\newcounter{KC}
\setcounter{KC}{0}

The four weakest matter criteria might be dubbed the \emph{kinetic criteria}. In order of logical strength:
\begin{description}
	\item[\refstepcounter{KC}\label{kc:change}Matter criterion strength \Alph{KC}:] The object under consideration is not constant/ static, but varies/ changes.\footnote{Within a model/possible world that is; we are not considering an inter-model/inter-world comparison. Note that we do not insist on the variation being across time.}
\end{description}

\begin{description}
	\item[\refstepcounter{KC}\label{kc:regular}Matter criterion strength \Alph{KC}:] The object under consideration changes in a regular fashion. 
\end{description}
\begin{description}
	\item[\refstepcounter{KC}\label{kc:diffeq}Matter criterion strength \Alph{KC}:] The object under consideration changes in a regular fashion that is describable by non-trivial differential equations. 
\end{description}
\begin{description}
	\item[\refstepcounter{KC}\label{kc:coupled}Matter criterion strength \Alph{KC}:] The object under consideration changes in a regular fashion. This change is (partially) in response to something external, and (thus) describable in terms of coupled differential equations that describe the coupling of the object to the external factor. % XXX Niels: This cange is necessary because the vaccuum Einstein equations are also coupled differential equations.
\end{description}

Newton's absolute space has none of these properties; Newtonian matter---such as planets and billiard balls---has all of them. Although the first criterion is thus in principle sufficient to separate space(time) and matter in the Newtonian framework (and also in special relativity), we will see below that Newton nevertheless had a much richer conception of matter. A Humean mosaic with mass values randomly sprinkled on top would be of exactly strength \hyperref[kc:change]{\makeAlph{\getrefnumber{kc:change}}}. Non-trivial sourceless\footnote{These are often referred to as vacuum solutions, but the notion of vacuum becomes ambiguous if the spacetime--matter distinction is ambiguous. What we are referring to is solutions with no other non-zero fields besides the metric.} solutions to GR, such as those containing gravitational waves, satisfy up to criterion \hyperref[kc:diffeq]{\makeAlph{\getrefnumber{kc:diffeq}}}. Generic metrics with sources that are solutions to GR satisfy all of the above criteria.

Criterion \hyperref[kc:coupled]{\makeAlph{\getrefnumber{kc:coupled}}} is a strong version of what is often called the action-reaction principle. Obeying a general action-reaction principle is logically speaking not stronger than criterion \hyperref[kc:regular]{\makeAlph{\getrefnumber{kc:regular}}}, if it is logically possible to regularly change due to something external without that implying that the change becomes describable in terms of differential equations. However, such a possibility is not relevant in the context of this paper. 

In the context of criterion \hyperref[kc:coupled]{\makeAlph{\getrefnumber{kc:coupled}}}, we could make the further distinction between objects whose dynamics is totally determined by external factors, and those which have their dynamics only partially determined externally, that is those with partially internal dynamics. A special case of demanding total determination is Einstein's 1918 version of Mach's principle: the demand that the metric field $\g$ be uniquely determined by the the mass-energy-momentum distribution of matter, $\T$ \cite{Einstein:1918}. This criterion was put under pressure in the debate between Einstein and  De Sitter following the discovery of the De Sitter solution to the Einstein field equations with cosmological constant in 1917 \cite{AEP8editorial,Hoefer:1994,Hoefer:1995}. We now know that the metric field in general relativity is merely constrained rather than determined by matter.\footnote{This becomes particularly clear from the decomposition of the Riemann curvature tensor into terms featuring only the Ricci tensor and terms featuring only the Weyl tensor. Only Ricci curvature is determined by the Einstein equations; the Weyl tensor encodes the free gravitational degrees of freedom---they are constrained but not fixed by the Einstein equations.} Hence, the set of solutions to the Einstein equations satisfying exactly up to criterion \hyperref[kc:diffeq]{\makeAlph{\getrefnumber{kc:diffeq}}} is the set of sourceless solutions; the same can be said for other field equations in which the gravitational field (in the case of GR the metric $\g$) has independent degrees of freedom. Having independent degrees of freedom is, however, not a necessary condition for being matter in any of these four senses.

\subsection{Energetic (matter) criteria} \label{energeticcrit}

The next three stronger criteria may be called the \emph{energetic criteria}. They require the object under consideration to \emph{carry energy}\footnote{We thus strongly disagree with Bunge, who equates energy to changeability and thereby conflates matter criteria \hyperref[kc:change]{\makeAlph{\getrefnumber{kc:change}}} through \hyperref[kc:tmunu]{\makeAlph{\getrefnumber{kc:tmunu}}} \cite{bunge2000}.}$^,$\footnote{Why is energy more relevant than, say, entropy? At least within thermodynamics it does not seem to be more special. It would be interesting to consider whether carrying entropy is in any way an indicator of being matter. If carrying entropy is considered necessary for being matter, one consequence would be that (the superfluid component of the two-fluid model of) perfect superfluids, such as those appearing in SFDM, would not count as matter. (The two-fluid model will be discussed in the second companion paper \cite{martenslehmkuhl2}.)} \cite{bunge2000}, a concept that can be cashed out in three distinct ways. In order of logical strength:

\begin{description}
	\item[Matter criterion strength E--G:] The object under consideration changes in a regular fashion. This change is (partially) in response to something external, and (thus) describable in terms of coupled differential equations, in such a way that the object can be said to carry exchangeable `energy' (and momentum)
	\begin{itemize} 
	\item[\refstepcounter{KC}\label{kc:energyregion}({\bf \Alph{KC}})] ascribable to a particular spacetime volume; 
	\item[\refstepcounter{KC}\label{kc:energypoint}({\bf \Alph{KC}})] ascribable to each point in spacetime;
	\item[\refstepcounter{KC}\label{kc:tmunu}({\bf \Alph{KC}})] representable by a stress-energy tensor, $T_{\mu\nu}$.
	\end{itemize}
\end{description}

Criterion \hyperref[kc:energyregion]{\makeAlph{\getrefnumber{kc:energyregion}}} insists on the energy being exchangeable, that is it `interacts' with some measurement apparatus whenever the required matter for such apparatus is present.\footnote{\label{apparatus} Here we assume that measurement apparatus must be made of matter. This assumption is of course questionable in light of the general thesis of this paper, and also, more specifically, in the context of attempts to reduce supposed matter to (or unify it with) spacetime.} After all, it might seem conceivable that an object has (non-exchangeable) energy, without satisfying criterion \hyperref[kc:coupled]{\makeAlph{\getrefnumber{kc:coupled}}}---being influencable by external objects---and even its converse---being able to influence external objects. To make it crystal clear that criterion \hyperref[kc:energyregion]{\makeAlph{\getrefnumber{kc:energyregion}}} is strictly stronger than \hyperref[kc:coupled]{\makeAlph{\getrefnumber{kc:coupled}}} we insist on the energy being exchangeable.\footnote{One might counter that this situation is not coherently conceivable, because such non-exchangeable energy would have no operational meaning and thereby not really any meaning whatsoever. Adding the qualifier `exchangeable' would thus be a pleonasm; no information is added. Such reasoning would open the door to finding the first three criteria meaningless as well. We will thus assume for now that the notions of change and energy still have some technical, if not operational, meaning in these situations. After all, as noted in \hyperref[apparatus]{fn.~\ref{apparatus}}, the notion of `operational' may have to be rethought anyway if the spacetime--matter distinction becomes ill-defined.}

Being kinetical (in any of the senses of criteria \hyperref[kc:change]{\makeAlph{\getrefnumber{kc:change}}}--\hyperref[kc:coupled]{\makeAlph{\getrefnumber{kc:coupled}}}) means that something is changing. Changing often means changing with respect to spacetime, in which case kinetic energy is indeed definable (relative to a frame), and if there is an action-reaction principle with (other) matter then this energy is exchangeable. Doesn't criterion \hyperref[kc:coupled]{\makeAlph{\getrefnumber{kc:coupled}}} imply \hyperref[kc:energyregion]{\makeAlph{\getrefnumber{kc:energyregion}}} and \hyperref[kc:energypoint]{\makeAlph{\getrefnumber{kc:energypoint}}} and perhaps even \hyperref[kc:tmunu]{\makeAlph{\getrefnumber{kc:tmunu}}}? Are they truly distinct---extensionally, or at the very least intensionally? Well, in GR the metric field, often taken to represent spacetime, is itself dynamical, which makes it less clear if and how energy is involved.  What we can definitely say, in the restricted context of GR, is that criterion \hyperref[kc:tmunu]{\makeAlph{\getrefnumber{kc:tmunu}}} is not satisfied by the metric, since gravitational energy is represented by a pseudo-tensor, rather than a tensor (and thus, \emph{a fortiori}, not by a stress-energy tensor $T_{\mu\nu}$). For Hoefer \cite{hoefer1996} and D\"urr \cite{duerr2019b,duerragainstread} this implies that criteria \hyperref[kc:energyregion]{\makeAlph{\getrefnumber{kc:energyregion}}} and \hyperref[kc:energypoint]{\makeAlph{\getrefnumber{kc:energypoint}}} are also not satisified by the metric. Hoefer believes that the concepts of carrying energy and energy being representable by a stress-energy tensor do not come apart\footnote{D\"{u}rr is more liberal, in that he allows all so-called `geometric objects' to potentially represent energy, even non-tensorial geometric objects. Pseudo-tensors are however not geometric objects in his sense \cite{duerragainstread}. } (hence the quotation marks around `energy' in criterion \hyperref[kc:energyregion]{\makeAlph{\getrefnumber{kc:energyregion}}} and \hyperref[kc:energypoint]{\makeAlph{\getrefnumber{kc:energypoint}}}):

\begin{quote}
I am inclined to believe ... that gravitational waves \emph{do not} in fact carry substantival energy content. ... [T]he fact that gravitational energy must be represented by a pseudo-tensor (and that no one pseudo-tensor has a privileged status for this representation) are the reasons for denying that gravitational energy is truly substantial.\cite[p.13; italics in original]{hoefer1996}\footnote{See also Hoefer's 2000 paper \cite{hoefer2000} and \cite{duerr2019a}.} %ik: dus, voor Hoefer, (en voor Patrick), the third and fourth criterion cannot come apart, or at least do not do so in GR. 
\end{quote}
This implication goes against a long tradition of taking gravitational waves (GWs) to carry energy (despite this energy not being represented by a tensor), whether at each point or only in each finite region. This tradition started with Einstein's first derivation of gravitational waves in the linearized limit \cite{Einstein:1916c, Einstein:1918b}, and was continued with Feynman and Bondi's sticky bead thought experiment designed to show that GWs can heat up matter \cite{ChapelHill:1957}. Subsequent advocates often invoke binary systems of which the orbital decay is claimed to be best explained by the system's energy and momentum being carried away by GWs. The tradition is perhaps most dramatically proclaimed by Rovelli, who does not only suggest that this carrying of energy puts the metric field on a par with the electromagnetic field, but that this shatters the spacetime--matter distinction altogether: 
\begin{quote}
	Let me put it pictorially. A strong burst of gravitational waves could come from the sky and knock down the rock of Gibraltar, precisely as a strong burst of electromagnetic radiation could. Why is the first ``matter'' and the second ``space''? Why should we regard the second burst as ontologically different from the second? Clearly the distinction can now be seen as ill-founded. \cite[p.193]{rovelli1997}
\end{quote}
Read argues, against Hoefer and D\"urr, that under certain assumptions and on a functional understanding of energy, there exist solutions to GR in which the metric can be attributed gravitational stress-energy (represented by a pseudo-tensor), albeit only in finite regions, not at a point (i.e., criterion \hyperref[kc:energyregion]{\makeAlph{\getrefnumber{kc:energyregion}}} but not \hyperref[kc:energypoint]{\makeAlph{\getrefnumber{kc:energypoint}}} is satisfied) \cite{read2018}.\footnote{The modern debate could profit from a more detailed study of the original debate on the nature of gravitational energy and the role of tensorial vs.\ pseudo-tensorial objects; see section II of Volume 7 and section VIII of Volume 8 of the Collected Papers of Albert Einstein for an overview, and \cite{Lehmkuhl:forthOUP} for analysis of and connections between the positions advocated by Einstein, Lorentz, Klein, Schr\"oedinger and Levi-Civita on these questions, as well as a discussion of how their insights could be used in the context of the modern debate.}

We will not attempt to settle this debate here. That the issues of GWs carrying energy and there being any substantial notion of gravitational energy in GR are controversial is sufficient reason to, at least \emph{prima facie}, separate criterion \hyperref[kc:energyregion]{\makeAlph{\getrefnumber{kc:energyregion}}} and \hyperref[kc:energypoint]{\makeAlph{\getrefnumber{kc:energypoint}}} from \hyperref[kc:tmunu]{\makeAlph{\getrefnumber{kc:tmunu}}}, and \hyperref[kc:coupled]{\makeAlph{\getrefnumber{kc:coupled}}} from \hyperref[kc:energyregion]{\makeAlph{\getrefnumber{kc:energyregion}}} and \hyperref[kc:energypoint]{\makeAlph{\getrefnumber{kc:energypoint}}}, and entertain the possibility that GR may contain solutions that satisfy up to criterion \hyperref[kc:coupled]{\makeAlph{\getrefnumber{kc:coupled}}} only as well as solutions that satisfy up to criterion \hyperref[kc:energyregion]{\makeAlph{\getrefnumber{kc:energyregion}}} or \hyperref[kc:energypoint]{\makeAlph{\getrefnumber{kc:energypoint}}}.

The final criterion perhaps does not deserve to be called a matter criterion, but it follows neatly in the hierarchy, and will become important when applying the matter criteria to SFDM:

\begin{description}
	\item[\refstepcounter{KC}\label{kc:mass}Matter criterion strength \Alph{KC}:] The object under consideration changes in a regular fashion. This change is (partially) in response to something external, and (thus) describable in terms of coupled differential equations, in such a way that the object can be said to carry exchangeable energy (and momentum) representable by a stress-energy tensor, $T_{\mu\nu}$, part of which is due to the rest mass of the object.\footnote{Note that this criterion is fulfilled by the relativistic fluids but not by electromagnetic fields. For the trace of the energy-momentum tensor of the former does not vanish (and thus allows for a rest frame and for the possession of rest mass), while the energy-momentum tensor of the latter vanishes: the continuum counterpart of the principle of the constancy of the speed of light.} In other words: the object is massive.
\end{description}

The tradition of considering `having mass' to be the litmus test for matter started with Newton in his \emph{Principia}, where he takes mass (arrived at via mass density times volume) to be the quantity of matter \cite[p.1]{newton1687} \cite{jammer2000}. This tradition was embraced by Maxwell, Kelvin, Tait and Clifford \cite{hoskins1915}, and was still influential in the 1920s when the aether was judged not to be matter exactly because it lacked rest mass. In Newtonian times such a criterion made perfect sense, as all well-understood (particle) matter had mass. In light of the current standard model of particle physics, and especially the existence of the massless photon,\footnote{In fact, many particles within the standard model are not \emph{always} massive; they only become massive via the Higgs mechanism once the electroweak symmetry spontaneously breaks in the early universe.} one may of course ask why mass would be so special. Why not consider spin, or isospin, or any other quantum number to be the essence of matter?\footnote{We would like to thank Radin Dardashti for this point.} Or, more generally, a disjunction of all of the above? Alternatively one may simply consider matter criterion \hyperref[kc:tmunu]{\makeAlph{\getrefnumber{kc:tmunu}}}, at most, to be sufficient for something to be matter; any stronger condition is asking too much. 

\section{Spacetime criteria} \label{S:geometric}

%\todo[inline]{Dennis removed loads of quotes in this section, compared to v8. Cite these authors (in the main text at the relevant criterion, or in a footnote.). DONE.  Also, mention spacetime functionalism: iels{Niels, give references. Wuthrich, Knox, Lam, Read and Menon, Le Bihan, Linnemann and Le Bihan, Baker}.}

The philosophical literature contains a second major family of criteria associated with the spacetime--matter distinction. We will refer to this family as the set of spacetime criteria. They are usually proferred as necessary and/or sufficient criteria for something to be spacetime or an aspect of spacetime structure in the subtantivalism--relationalism debate.

Both substantivalism and relationalism are realist positions about space(time), they merely differ on the relative fundamentality of space(time) and spati(otempor)al relations between material objects. Within the context of the modern substantivalism--relationalism debate one needs to consider two issues: 

\begin{enumerate}

\item[i)] the core issue: whether spacetime and matter are separate fundamental entities (substances) or whether spacetime is derivable from matter in some sense (for example from the relationships between material bodies). This part of the debate is the direct continuation of the debate between Leibniz and Clarke in the context of pre-relativistic mechanics. 

\item[ii)] the prior issue of fixing the referents of the terms in the core issue, in particular what the referent of the term `spacetime' is in GR. The main candidates in the philosophical literature on the substantivalism--relationalism debate are a) the manifold $M$ by itself, b) the manifold $M$ and the metric field $\g$ together, c) the metric field $\g$ by itself \cite{earman1987,maudlin1993,hoefer1996}.

\end{enumerate}

Only the second issue matters to us in the context of this paper, or rather the connected question of what is a lower bound, a minimal set of properties for \emph{something} to count as spacetime or as an aspect of spacetime structure. In the following we will consider some candidate criteria for calling something a spacetime. It will be important to keep two things in mind.\footnote{Both of these caveats cohere with Baker's understanding of the notion of spacetime as a cluster concept \cite{bakerstfunc}, with which we sympathise to some extent. A cluster concept is a way of making precise how certain properties can count toward the application of the concept being appropriate. Although these properties are jointly sufficient for the concept, say spacetime, to apply, they are not jointly nor even individually necessary---but, in order to avoid triviality, they are disjunctively necessary in the sense that at least some of the properties must obtain in order for the concept to apply. (Note that understanding spacetime as such a cluster concept, rather than insisting on a set of criteria that is both jointly necessary and sufficient, immediately suggests that we should not expect every (logically possible) object to fall into exactly one of two categories, spacetime or matter \cite[\S 7]{martenslehmkuhl2}.) This paper does not intend to give an exhaustive list of properties that might count toward the concept of spacetime applying. It merely aims to include one set of criteria that is jointly sufficient for something being spacetime---a set of criteria that applies to $\Phi$ in particular (as well as to the metric in GR).} 

Firstly, the list of three structures discussed in the philosophical literature (issue ii) is too limited for the space of theories we envisage, although this is perhaps not surprising given that the philosophical debate presupposed GR. The physics literature considers a much larger variety of possibilities, see for instance Sharpe \cite{sharpe1997}, Vizgin \cite{vizgin1994}, Goenner \cite{goenner2004} and Blagojevi\'{c} \& Hehl \cite{blagojevichehl2013}. In the following, we will go slightly beyond these three structures, but will still fall short of considering the full range. Instead, we will stay close to general relativity and consider one set of (many possible sets of) criteria that are jointly sufficient for something to be spacetime, namely the set that is relevant to the case study at hand. Note that we are thus not claiming that these criteria are necessary for being spacetime\footnote{For instance, neither superspace \cite{menonforth} nor discrete `spacetimes' satisfy these criteria, but we do not believe that this suffices to rule them out as spacetimes.}---as in fact we do not think they are, although we do believe that many of these criteria are typically fulfilled by spacetimes. Defining spacetime exhaustively in the context of the full space of theories will require going beyond this set of criteria.

Secondly, although some of these criteria are logically stronger or weaker than others, others are logically independent of each other. Hence, we will not label them alphabetically as we did to indicate the total logical ordering of the matter criteria.

\subsection{Mathematical (spacetime) criteria}

\newcounter{GC}
\setcounter{GC}{0}

A first subset of the spacetime criteria might be called the \emph{structural} \cite{lehmkuhlthesis} or \emph{abstract} or \emph{mathematical criteria}. A first candidate is as follows:

\begin{description}
	\item[\refstepcounter{GC}\label{gc:mani}Spacetime criterion M:] The object under consideration is (faithfully representable by) a differentiable manifold, i.e.\ a (topological) manifold plus differentiable structure.
\end{description}

Leibnizian spacetime \cite[p.30--31]{earman1989} satisfies this criterion. However, Newton's bucket experiment is designed to show that we need a way of distinguishing straight motion from curved motion\footnote{For simplicity's sake, we are ignoring that strictly speaking one only needs a standard of rotation, not a full standard of curved motion \cite{saunders2013,knox2014,wallace2017}.}---it is exactly for this reason that absolute space has empirical consequences, according to Newton. As any affine connection grounds a distinction between geodesics and  non-geodesics, this suggests adding at least one affine connection to obtain the following stronger criterion: 

\begin{description}
	\item[\refstepcounter{GC}\label{gc:conn}Spacetime criterion M$\nabla$:] The object under consideration is (faithfully representable by) a (differentiable) manifold endowed with an affine connection.
\end{description}

But what about physical theories where a distinction between geodesic and non-geodesic paths is not empirically significant, say theories for which Leibnizian space(time) would be a sufficiently rich space(time)?  

Moreover, objects that satisfy either of these two criteria (and feature in a diffeomorphism invariant theory) seem to run up against the hole argument.\footnote{The manifold substantivalism that Earman and Norton target uses criterion M explicitly, but their argument would work equally well against a form of substantivalism that adds a connection.} Of course, one might attempt to avoid the hole argument using one of several responses in the literature. For instance, it may be pointed out that Earman and Norton assume primitive/fundamental transworld identities of the spacetime points, and that the hole argument dissolves if one strips these identities away \cite{Butterfield:1989, pooley2002b,pooleybookforthcoming,pooley2013c}. The objects that remain would still satisfy criterion M or M$\nabla$, and therefore deserve the name `spacetime'. Moreover, even if the hole argument would go through, that would not imply that these two criteria are bad criteria for denoting something as spacetime. For instance, in the context of GR, it would merely imply that GR is best interpreted relationally, and therefore that spacetime is not fundamental---albeit still real.\footnote{Of course, such a non-fundamental spacetime would not feature fundamental transworld identities either---but see Maunu \cite{maunu2005}.} The hole argument is thus not a reason against taking either of these two conditions as criteria for being spacetime.

Some considerations that do pose a real problem turn out to arise even prior to the introduction of matter dynamics,\footnote{cf.~\cite[\S 9.3.2]{lehmkuhlthesis}} as was the case for the bucket and hole arguments. Firstly, configuration spaces or phase spaces could also satisfy the above two criteria. How to distinguish them from the type of space that we are after? Secondly, these criteria do not distinguish n+1 dimensional space from n+1 dimensional spacetime. Let us consider the following stronger criterion:

\begin{description}
	\item[\refstepcounter{GC}\label{gc:conn+metric}Spacetime criterion M$\nabla$g:] The object under consideration is (faithfully representable by) a (differentiable) manifold with an affine connection, plus a Lorentzian metric field\footnote{For a more nuanced position, see Dewar \& Eisenthal \cite{dewareisenthalms}, who argue for a middle way between bare-manifold and manifold-plus-metric accounts of spacetime.}$^,$\footnote{Of course the topological and differentiable structure and \emph{a fortiori} the connection mentioned in the first two spacetime criteria also require a metric in order to be defined, namely a metric on the space of coordinates, but that metric is in general not the same metric as the metric field living on the manifold. For instance, in GR the topology is induced from a Euclidean metric on the space of coordinates, but has a Lorentzian metric living on the manifold. We are grateful to Tushar Menon for pointing this out.} on that manifold (which may or may not be compatible with the connection).
\end{description}
This criterion is too strong, or rather something weaker will do the job of distinguishing between two types of dimensions---and thus between space and spacetime:

\begin{description}
	\item[\refstepcounter{GC}\label{gc:conn+lrntzstruct}Spacetime criterion M$\nabla$C:] The object under consideration is (faithfully representable by) a (differentiable) manifold with an affine connection, plus a Lorentzian conformal structure, i.e.\ an equivalence class of Lorentzian metrics, on that manifold.%\todo{Niels, reinstate the footnote on Kaluza-Klein and Maudlin?}
\end{description}

Conformal structure suffices to distinguish between space and spacetime---it distinguishes two types of dimensions, one usually referred to as temporal, the other as spatial---while affine structure suffices to distinguish between geodesic and non-geodesic paths through spacetime.
 
One problem that remains is that we do not know how to distinguish an n+1 spacetime from a 1+n spacetime. For instance, why do we take fourdimensional spacetimes to have one temporal dimension and three spatial dimensions, rather than the other way round? Is it simply essential to time that it is a single dimension, and essential to space that it may have multiple dimensions? %\todo{Niels, reinstate the footnote on Pooley and Brown?} 
But where would this leave 1+1 spacetimes? Or is the distinction between space and time less relevant in such a context---after all, GR in two dimensions exhibits a cornucopia of pecularities \cite{fletcher2018}?

A further problem arises. Reference to \emph{the} metric (or conformal structure) in some of the above criteria is ambiguous, as there may be more than one. For one thing, one can always just define extra metrics, such as Finsler metrics. One might respond that the above criteria implicitly refer to the metrics that are fundamental within the theory under consideration, that is metrics that are postulated as part of the fundamental ontology of the theory. This would however rule out, by fiat, mainstream positions such as relationalism---which is a realist but non-fundamentalist position about spacetime---as well as the dynamical approach to special relativity \cite{brown2005}. It would also be very much against the spirit of spacetime functionalism\todo{This still needs to be defined somewhere in the paper.}. Even if we were to take this route, we would expect the problem of multiple metrics to resurface in some theories of modified gravity/spacetime---where we might expect the modification to generate a second effective metric---and explicitly in bimetric theories, where both metrics are part of the fundamental ontology. Application of spacetime criterion M$\nabla$g (and in some cases also criterion M$\nabla$C) would then suggest that this theory exhibits two spacetimes, which both share the same differentiable manifold, but consist of a different metric. It seems a sensible restriction on the concept of spacetime that there can be at most one spacetime in each model of a theory at each level of description. Worries similar to those about multiple metrics apply to theories with multiple affine connections, and even in theories with just one metric and one affine connection if those are not compatible. In the latter case, it is ambiguous whether a trajectory (say of a test particle) is a geodesic, as it may be an affine geodesic but not a  metric geodesic, or vice versa. 

Furthermore, we still have not managed to distinguish configuration and phase spaces from the space(time) that we are after. (Imposing that the metric (or conformal structure) be dynamic would certainly rule out configuration space and phase space, but at the cost of denying Special Relativity the status of a spacetime theory.) The reason is that the criteria so far talk only of abstract, mathematical objects but have not made any connection with physics, that is with the (observable) behaviour of matter \cite[p.12]{hoefer1996} \cite[esp.~p.89]{lehmkuhl2008}.\footnote{One may thus wish to call the spacetime candidates in this subsection `theoretical spacetimes' and such spacetimes that furthermore satisfy the physical criteria in the next subsection the `physical spacetimes' or `operational spacetimes'. It is however important to distinguish this usage from the usage by Read and Menon, for whom both the concepts of theoretical and operational spacetime refer to the dynamics of matter \cite{readmenonforth}.} Perhaps by moving from mathematics to physics we can thereby also resolve the impasse arising from spacetime and configuration space being representable by the same type of mathematical object.

\subsection{Physical (spacetime) criteria}

We have come to the second and final subset of the spacetime criteria, the \emph{physical} criteria. `Physical' is meant to oppose `merely mathematical', not to be synonymous to `material', a mistaken equivocation that often occurs in the context of the matter criteria (cf.\ fn.\ref{physicalvsmatter}). These criteria take into account the connection between spacetime and matter. The first advantage that this connection provides is that it allows us to distinguish between n+1 and 1+n spacetimes:%\todo{YOu said on one of the previous pages that this was already taken care of by criterion \ref{gc:compatible}. Edit: is this todo note irrelevant now the compatibility condition has been removed?? Check this.} 

\begin{description}
	\item[\refstepcounter{GC}\label{gc:ivp}Cauchy (spacetime) criterion:] A Lorentzian signature separates the dimensions into two types. The dimension(s) with respect to which a well-defined initial value problem can be formulated---for the matter `living in' that spacetime and the dynamics governing that matter---form(s) the temporal type; the other dimensions are of the spatial type. \cite{callender2017}
\end{description}

In practice, this ensures that there is typically one temporal and $n$ spatial dimensions, for multiple time dimensions would make a well-defined initial value problem all but impossible (\emph{pace} Weinstein \cite{weinstein2008many}).

Two further criteria concern the connection between spacetime and matter. In response to the question of what reasons we have to believe that GR can be interpreted as geometrising the gravitational field---which we take to be equivalent to gravity becoming part of spacetime in GR---one might give a two-part answer.\footnote{See \cite[Chapter 9]{lehmkuhlthesis} for details, and the inspiration for the weak geodesic criterion below, applied to the question of whether the gravitational field in GR, as well as both the gravitational and electromagnetic field in Weyl's theory and in Kaluza-Klein theories, can be interpreted as an aspect of spacetime structure.} First, in GR we can describe gravity solely in terms of the structure of spacetime.\footnote{One might not describe geometrised forces in terms of curvature (only), but in terms of other structures, such as metric and affine properties, and those deriving from them, like torsion and con-torsion, or even topological ones \cite{wheeler1962,lehmkuhl2018}.} Secondly, in GR gravitationally charged (i.e.\ massive) test particles move on the timelike geodesics of the connection (which is compatible with the metric) even in the presence of gravity, and lights rays move on the null geodesics of the same connection.\footnote{Note that Einstein opposed the idea that any of this means that gravity is reducible to the geometry of spacetime; see Lehmkuhl \cite{lehmkuhl2014} for details.}  Our second physical criterion is thus:

\begin{description}
	\item[\refstepcounter{GC}\label{gc:geodesic}Strong geodesic (spacetime) criterion:] For any choice of initial conditions, if test particles (and massive bodies that can be idealised as test particles) were around, then they would all follow the time-like geodesics of the same affine connection or metric whose null geodesics light rays would follow if they were around. 
\end{description}

It is often simply assumed that in GR the Levi-Civita connection fulfils this criterion.\footnote{See e.g. \cite{Malament:2012}, p.120-121.} And because this connection also gives rise to the curvature tensor that features on the left-hand side of the gravitational field equations, and is compatible to the metric tensor that solves these field equations, it is possible to interpret GR as showing that the gravitational potential $g_{\mu \nu}$ is part of spacetime structure. If one does not want to assume that that test particles move on timelike geodesics, then different geodesic theorems (one of which we discuss below) allow one to derive that they do. When it comes to deriving the assertion that light rays move on null geodesics, then the choice between possible ways of deriving this is much more limited; the most common argument procceds via the geometrical-optical limit of GR, which is far from having the status of a theorem.\footnote{Note, however, that Geroch and Weatherall's approach \cite{geroch2018motion} towards deriving geodesic motion, which relies on a new mathematical concept they call ``tracking'', allows to derive the assertion that bodies constructed from wave packets of Maxwell fields `track' null geodesics.  They argue (p.626) that this result `reflects' the result normally gained via the geometric-optical limit, namely that light rays move on null geodesics, in a precise way.} 

This might lead one to demand a weakened version of the geodesic criterion, one that remains agnostic on whether light rays do indeed move on null geodesics:

\begin{description}
	\item[\refstepcounter{GC}\label{gc:geodesicweak} Weak geodesic (spacetime) criterion:] For any choice of initial conditions, if test particles (and massive bodies that can be idealised as test particles) were around, then they would all follow the geodesics of the same affine connection or metric. 
\end{description}
For this criterion the distinction between timelike and null is not relevant anymore.

But note that though this criterion is weaker, we are here searching not for necessary conditions for something to be a spacetime or part of spacetime structure; the array of generalizations of pseudo-Riemannian geometry that arose after GR is far too rich for that, and each such generalization, with its multiple curvature tensors, torsion, contorsion and non-metricity tensors, is in principle a candidate for a way spacetime could be.\footnote{See \cite{goenner2004}, section 2.1, for an overview of such structures, many of which were tried out as candidate spacetimes in the context of unified field theories.} For our purposes, designed to tell whether a newly introduced field in superfluid dark matter theory counts as part of spacetime structure, a sufficient condition is enough. And here the stronger condition, the strong geodesic criterion, seems to provide for the more cautious path. 

One tentative possibility for a jointly sufficient (albeit not necessary) set of criteria for something to be interpretable as spacetime would then be the following: it must adhere to spacetime criterion M$\nabla$C, such that the Cauchy criterion is satisfied and test particles and light rays follow the geodesics of its metric or connection. This solves our problems of multiple metrics, multiple connections and incompatible connections and metrics: the unique object whose geodesics are followed by test particles and light rays is the object that is part of spacetime. If a theory modifies spacetime/gravity in such a way that test particles and light rays would follow the geodesics of this novel, modified, effective metric, one may say that particles moving `under the influence' of this modification are forcefree. Thus, when there is a candidate modified/effective metric, we can determine whether this is the physically relevant metric by checking whether test particles and light rays would follow the geodesics of this modified object (or rather the geodesics of the connection compatible with that new metric) rather than those of the standard metric posited by the theory.

However, the above proposal is incomplete, as a final important physical spacetime criterion remains, which we might call the chronogeometricity criterion or the Strong Equivalence Principle \cite{readbrown2018}. We need not only ensure that test particles and light rays survey (the geodesics of) the metric or connection, but also that the (luminous) matter from which one may construct rods and clocks couples to the (same) metric or connection in the right way. That is, there should be no (non-negligible) curvature terms appearing in the dynamical equations which would prevent those rods and clocks from functioning properly, i.e.\ from behaving, locally, as they would in special relativity: exhibiting time dilation, length contraction, etc.

\begin{description}
	\item[\refstepcounter{GC}\label{gc:sep}Chronogeometricity (spacetime) criterion:] The Strong Equivalence Principle is satisfied, i.e.\ Special Relativity is locally valid, i.e.\ there are no non-negligible curvature terms in the local equations of motion.\footnote{For more precise definitions, see \cite{readbrown2018,lehmkuhlequivalenceprinciples}.}
\end{description}

A more demanding set of jointly sufficient criteria for an object to be interpretable as part of spacetime structure, and the one we will work with in the following, is then that the object is defined on a manifold fulfilling spacetime criterion M$\nabla$C, such that the `constraints' on matter `living in' that spacetime, as described by the Cauchy, strong geodesic and chronogeometricity criteria, are satisfied.\footnote{Arguably there is one additional relevant criterion: one needs independent reasons to believe that the dimensionality of the manifold matches the dimensionality of physical spacetime. Without this additional constraint, the 5-dimensional metric of a Kaluza-Klein theory would satisfy all the criteria in the main text, even though it is arguably the 4-dimensional metric that one obtains from projecting the 5-dimensioal metric into four dimensions that corresponds to physical space \cite{lehmkuhlthesis}.} Note that if all of these conditions are imposed together, then the metric and curvature tensor referred to in the chronometricity criterion need to be those compatible with the metric referred to in the geodesic criteria. If there were no further geometric structures defined on the manifold, then this would imply that the conformal and affine structure referred to in criterion M$\nabla$C would be compatible with one another. This in turn means that the part of spacetime structure that is surveyed by rods, clocks, test particles and light rays needs to be either (effectively) pseudo-Riemannian or Weylian.\footnote{See \cite{EhlersPiraniSchild:1972} for a proof to that effect.} Of course, that seems to restrict the set of allowed spacetime structures at first sight. But note again that we are not saying that all these conditions \emph{have to be} fullfilled for something to be a spacetime or an aspect of spacetime structure; we only say that \emph{if} they are fullfilled, \emph{then} the object in question can be \emph{interpreted} as part of spacetime structure.   

With the matter and spacetime criteria in hand, it is time to turn to our case study, superfluid dark matter theory (SFDM), and see how the newly introduced field of the theory fares in light of our criteria---and how the criteria fare in light of the physics. 

\section{`Superfluid Dark Matter theory' (SFDM)} \label{FDMintro}

Theories labeled as dark matter theories have traditionally done well at the level of cosmology and galaxy clusters, but less so at the level of galaxies. The opposite is the case for theories labeled as modifications of gravity and/or spacetime. A promising approach to breaking this stalemate is to find a single novel entity for which there is a natural, physical, dynamical reason why it behaves like DM on large scales and like MG on galactic scales. In this respect it is relevant to note that to mediate a long-range force in galaxies, a massless messenger (force carrier) is needed. A natural candidate presents itself: the quantised soundwaves of a superfluid, i.e.\ phonons, which are Goldstone bosons and thus massless. In the Standard Model of Particle Physics, matter (in the broad sense used in this paper) is divided into bosonic force carriers and fermionic matter (in a narrower sense of matter).\footnote{We are excluding the Higgs boson here, since, although it is a boson, it is not associated with one of the four fundamental forces.} But there is no reason why matter, in this narrow sense, could not also be bosonic, and it is not uncommon for new dark matter theories to postulate a bosonic dark matter field. If the associated particles self-interact (repulsively), they can form a superfluid Bose-Einstein condensate (BEC), which carries phonons. In other words, in this phase one cannot associate with the field a set of individual (nearly) collisionless particles; it is best described in terms of collective excitations. If these phonons cohere they can mediate a long-range force. If the phonons are described by the appropriate Lagrangian, they mediate a MONDian force. In order for the superfluid BEC phase to obtain, the De-Broglie wavelength of the $\Phi$-particles needs to be larger than the mean interparticle separation \cite[p.5]{berezhiani2016} \cite[\S 2.2]{annett2004} (or equivalently, the associated temperature needs to be below a critical value). In order for the phonons to cohere the BEC needs to be allowed sufficient time to thermalise \cite[p.5-6]{berezhiani2016}. If parameters are chosen such that those two conditions obtain only in galaxies, but not at larger scales, this would provide a natural reason for why the MONDian behaviour of the single novel entity appears only in galaxies. This approach to resolving the dark phenomenon stalemate is created and developed by Berezhiani and Khoury, under the name `Superfluid Dark Matter Theory'. We will outline the original version of their theory \cite{berezhiani2015,berezhiani2016} in this section.

%The theory takes General Relativity with a cosmological constant and `normal' matter content as its starting point, i.e. the Lagrangian is \todo{This is the incorrect Lagrangian. The whole point is that the matter Lagrangian is a function of the physical metric, not of the Einstein metric. This is the only and crucial fact that will determine our 'evaluation of Phi according to the spacetime criteria'.}

\todo{IGNORE: Berezhiani 2016 p17: a0 depends on velocity dispersion... on cosmological scales it will be different form the value that MOND people take it to be.}

This theory postulates a self-interacting, massive, complex scalar field $\Phi$ with global $U(1)$ symmetry, via the following term:
\begin{equation} \label{eqBerezhiani}
	\mathcal{L}_{\Phi} = - \frac{1}{2} \left( |\partial_{\mu} \Phi|^2 + m^2 |\Phi|^2 \right) - \frac{\Lambda^4}{6(\Lambda_c^2+|\Phi|^2)^6} \left( |\partial_{\mu}\Phi|^2 + m^2|\Phi|^2 \right)^3
\end{equation}
with $\Lambda_c$ and $\Lambda$ two constants introducing two mass/energy scales.\footnote{$\Lambda_c$ ensures that the theory admits a $\Phi=0$ vacuum \cite[p.15]{berezhiani2016}.} (The main justification for this choice of Lagrangian comes from reverse engineering: it is this Lagrangian that will ultimately reproduce MOND in the Galactic regime, as discussed below.) 

Spontaneous symmetry breaking of the global $U(1)$ symmetry yields, in the non-relativistic regime, the following Lagrangian for the associated Goldstone bosons---these massless phonons being represented by the scalar field $\theta$, the phase of $\Phi = \rho e^{i(\theta + mt)}$:\todo{IGNORE: Berezhiani 2017 p5: the interaction term makes means that we have a pseudo-superfluid, and the phonons will in fact acquire a mass via radiative corrections, even though this has no observable effects on galactic scales.}

\begin{equation} \label{eqX}
	\mathcal{L}_{T=0,\neg rel,\theta} = \frac{2\Lambda(2m)^{3/2}}{3}X\sqrt{|X|},
\end{equation}
with 
\begin{equation}  \label{eqX2}
	 X \equiv \dot{\theta} - m\Phi - \frac{(\vec{\nabla}\theta)^2}{2m},
\end{equation} 
where $\Phi$ is now interpreted as the external gravitational potential. (This identification receives its justification when one derives this non-relativistic description from the relativistic description, i.e.\ \hyperref[crucial]{Eq.(\ref{crucial})}, with $\tilde{g} = \tilde{g}^{SFDM}$ as given by \hyperref[gFDM]{Eq.(\ref{gFDM})}.) To lowest order in the derivatives, superfluid phonons are in general described  by a scalar field $\theta$ governed by a Lagrangian $\mathcal{L} = P(X)$, with $X$ given by \hyperref[eqX2]{Eq.(\ref{eqX2})} \cite[p.3]{berezhiani2016}\cite{son2002}. The specific choice of $P$ given by \hyperref[eqX]{Eq.(\ref{eqX})} uniquely determines the specific type of superfluid, namely one that interacts primarily through three-body interactions, i.e.\ with an equation of state $P \propto \rho^3$.

The interaction between the phonons and the regular (i.e.\ luminous) matter fields is then added to this Lagrangian as an ``empirical term'' \cite[p.8]{berezhiani2016} (as opposed to being derived from an interaction term in the fundamental Lagrangian). In the relativistic regime we may describe this via the metric $\tilde{g}_{\mu\nu}$, sometimes referred to as the physical metric---whether such metrics deserve this name will be discussed in \hyperref[geoFDM]{\S \ref{geoFDM}}---because all luminous-matter fields couple universally to this metric (and to the Einstein metric $g_{\mu\nu}$ only indirectly, via this physical metric):\footnote{\label{newFDM} In a later paper \cite{berezhiani2017}, photons are not coupled to the effective metric, but to the Einstein metric. Some conclusions reached in \hyperref[geoFDM]{\S \ref{geoFDM}} do not apply to that alternative version of the SFDM theory. (See also the end of Section 7 of the sequel paper \cite{martenslehmkuhl2}.)}
\begin{equation} \label{crucial}
\mathcal{L}_{\ rel,\ int} = \mathcal{L}(\tilde{g}_{\mu\nu},\psi^{\alpha},\psi^{\alpha}_{;\mu | \tilde{g}})
\end{equation}
with $\psi^{\alpha}$ the luminous-matter fields, and $\psi^{\alpha}_{;\mu | \tilde{g}}$ denoting the covariant derivative with respect to $\tilde{g}$. The physical metric of SFDM is inspired by that of TeVeS. 

TeVeS, a theory usually referred to as a modification of gravity, postulates two new dynamical fields---a real scalar field $\phi$ and a 4-vector field $A_{\mu}$---which, together with the Einstein metric $g_{\mu\nu}$, constitute the effective metric $\tilde{g}_{\mu\nu}^{TVS}$. This physical metric is disformally related to the Einstein metric \cite{bekenstein2004}\cite[\S 3.4]{clifton2012}.  That is, it does not stretch the Einstein metric equally in all directions, which would leave all angles and thus shapes invariant, as with a conformal transformation, but instead \emph{stretches} it by a factor of $e^{-2\phi}$ in the directions orthogonal to $A^{\mu} \equiv g^{\mu\nu}A_{\nu}$ while \emph{shrinking} it by the same factor in the direction parallel to $A^{\mu}$, where the Sanders 4-vector field $A_{\mu}$ is unit-time-like with respect to the Einstein metric ($g^{\mu\nu}A_{\mu}A_{\nu} = -1$) \cite{bekenstein1993,bekenstein2004,clifton2012}:
\begin{equation} \label{tevesmetric}
\begin{array}{rcl}
\tilde{g}^{TVS}_{\mu\nu} & = & e^{-\frac{2\alpha\Lambda}{M_{Pl}}\phi}(g_{\mu\nu}+A_{\mu}A_{\nu}) - e^{\frac{2\alpha\Lambda}{M_{Pl}}\phi}A_{\mu}A_{\nu} \\
& = & e^{-\frac{2\alpha\Lambda}{M_{Pl}}\phi}g_{\mu\nu}-2A_{\mu}A_{\nu}\text{sinh}(\frac{2\alpha\Lambda}{M_{Pl}}\phi) \\
& \approx & g_{\mu\nu} - \frac{2\alpha\Lambda}{M_{Pl}}\phi(g_{\mu\nu}+2A_{\mu}A_{\nu}),
\end{array}
\end{equation}
with $M_{Pl}$ the Planck mass and $\alpha$ a dimensionless coupling constant.

SFDM modifies TeVeS in two ways: one semantic and one syntactic modification. The semantic revision is to not add yet another (vector) field, but to identify the four-vector $A_{\mu}$ with the unit four-velocity $u_{\mu}$ of the (normal fluid component \cite[\S 5.3.3]{martenslehmkuhl2} of the) previously added scalar field $\Phi$. Syntactically, the TeVeS factor of 2 is generalised to obtain: 
\begin{equation} \label{gFDM}
	\tilde{g}_{\mu \nu}^{SFDM} \approx g_{\mu \nu} - \frac{2\alpha\Lambda}{M_{Pl}}\theta \left( \gamma g_{\mu \nu} + (1+\gamma)u_{\mu}u_{\nu} \right),
\end{equation} 
with TeVeS being recovered for $\gamma = 1$ (and a metric conformally related to $g_{\mu \nu}$ for $\gamma = -1$). 

The full relativistic form may be used to calculate gravitational lensing effects. \todo{IGNORE: This doesnt' seem true anymore in berezhiani2017. See handwritten notes at the end of october 2018 on important points in berezhiani 2016 and 2017.} In the context of recovering the galaxy rotation curve component of the dark phenomena we only require the non-relativistic approximation:
\begin{equation} \label{eqTeVeSlike}
	\mathcal{L}_{\neg rel,int} = - \alpha \frac{\Lambda}{M_{Pl}}\theta \rho_b,
\end{equation}
where $\rho_b$ is the baryon density. The total effective, non-relativistic Lagrangian can then be shown to reproduce \cite[p.10-11]{berezhiani2016}, under suitable approximations (such as $\theta$ being static and spherically symmetric), the MONDian result $a_{MOND} = \sqrt{a_0 \frac{GMb(r)}{r^2}}$ for a baryonic particle with mass $M_b$, after identification of $a_0 = \frac{\alpha^3\Lambda^2}{M_{Pl}}$.

To a particle physicist, the fractional power of $X$ (\hyperref[eqX]{Eq.\ref{eqX}}), 3/2, although required if one aims to eventually regain MONDian behaviour \cite[\S IV]{berezhiani2016}, might seem strange---it is, for instance, less straightforward to draw corresponding Feynman diagrams. In condensed matter theory, such powers are far from rare. As mentioned, this specific fractional power corresponds to a phonon superfluid, with equation of state $P \propto \rho^3$.

Superfluidity only occurs at sufficiently low temperature. This naturally distinguishes between galaxies and galaxy clusters.  Due to the smaller velocities in galaxies,\footnote{Given a mass $m$ and density $\rho$ \cite[Eqs.\ 8 \& 80]{berezhiani2016}.} the superfluid description is appropriate there,\footnote{Since the local phonon gradient induced by the Sun is too large to satisfy the criteria for a superfluid Bose-Einstein condensate, the condensate loses its coherence, which allows SFDM to avoid solar system constraints \cite[\S V]{berezhiani2016}.} exactly where MOND is successful. In the clusters the velocity/ temperature is too high, and one finds either a mixture of the superfluid and normal phase, or only the normal phase, suggesting that the theory might exemplify the usual successes of dark matter theories at that level.

%It is not totally clear from the original FDM papers how the above Lagrangians are to be pieced together. ............................................... Just take the Phi lagrangian plus the effective metric lagrangian. Interpreting theta as phonons (with u its associated unit four-velocity) would mean that the effective metric reduces to the Einstein metric when there are no phonons around. In that case we would retrieve GR plus an extra scalar field, making it plausible that the success of lambdaCDM are reproduced in that scenario. However, technically speaking theta is defined via ....... and hence can be nonzero even when the temperature is not below the critical temperature. Do we need to make the physical stipulation that the L-effectivemetric only applies when we are below the critical temperature? But how does that follow from the mathematics? 

%......................(To what Lagrangian this new term is added is controversial. The obvious suggestion would be the Einstein-Hilbert Lagrangian plus some Lagrangian for luminous matter, but we will discuss in \hyperref[discussion]{\S \ref{discussion}} that this cannot be correct, or at least not the complete story.)

\section{Is SFDM a Theory of Dark Matter or of Modified Gravity/Spacetime?} \label{S:FDMeval}

%\todo{Alternative title: what's the matter with spacetime?}

Having introduced the two families of matter and spacetime criteria, as well as SFDM with its novel complex scalar field $\Phi$, we are finally in the position to ask the main question: which label(s)---dark matter or modified gravity/spacetime, or both, or neither---does this new field $\Phi$ within SFDM deserve? In this section we first evaluate $\Phi$ with respect to the matter criteria (\hyperref[kineval]{Subsection~\ref{kineval}}) and subsequently with respect to the spacetime criteria (\hyperref[geoFDM]{Subsection~\ref{geoFDM}}).

\subsection{Matter evaluation of SFDM} \label{kineval}

\todo{IGNORE: Niels, note that the phonons should be massless, if they are indeed Goldstone bosons. It is Phi that is massive. It is also that mass that causes a gravitatoinal interaction, according to Berezhiani et al. Edit: Berezhiani 2017 p5: the interaction term makes means that we have a pseudo-superfluid, and the phonons will in fact acquire a mass via radiative corrections, even though this has no observable effects on galactic scales. Niels, quote/cite this! }

\begin{table}
		\begin{tabular}{lcc}
		\bf \sc Object & \bf \sc Theory & \bf \sc Matter strength \\ 
		\hline 
		Newtonian spacetime & Newtonian Gravity & -- (static) \\
		Minkowski spacetime & Special Relativity & -- (static) \\
		Non-trivial $g_{\mu\nu}$ & General Relativity & \hyperref[kc:coupled]{\makeAlph{\getrefnumber{kc:coupled}}} (action-reaction) or   \\
		 & & \hyperref[kc:energyregion]{\makeAlph{\getrefnumber{kc:energyregion}}} (E in region) \\
		Photon & Standard Model  & \hyperref[kc:tmunu]{\makeAlph{\getrefnumber{kc:tmunu}}} ($T_{\mu\nu}$) \\
		Electron & Standard Model & \hyperref[kc:mass]{\makeAlph{\getrefnumber{kc:mass}}} (mass) \\
		Scalar field $\Phi$ & Superfluid Dark Matter  & \hyperref[kc:mass]{\makeAlph{\getrefnumber{kc:mass}}} (mass) (or \hyperref[kc:tmunu]{\makeAlph{\getrefnumber{kc:tmunu}}} ($T_{\mu\nu}$)) \\
		\end{tabular}
		\caption{Application of the matter criteria to the case study and to other familiar objects that provide a contrast.} 
		\label{kineticeval}
\end{table}

The matter criteria have been put forward, in their various strengths, as necessary and/or sufficient conditions for something to be matter. Before using them to evaluate $\Phi$, let us briefly rank some familiar objects, in order to eventually provide contrast with $\Phi$. Newtonian spacetime and Minkowski spacetime satisfy none of the matter criteria, and would thus, as expected, be considered not to be matter (\hyperref[kineticeval]{Table~\ref{kineticeval}}). A paradigmatic matter field such as that of the electron in Quantum Electrodynamics satisfies all matter criteria up to strength \hyperref[kc:mass]{\makeAlph{\getrefnumber{kc:mass}}}. A photon satisfies up to exactly criterion \hyperref[kc:tmunu]{\makeAlph{\getrefnumber{kc:tmunu}}}. Remember that we are in the business of distinguishing matter from spacetime, not stable (fermionic) matter from (bosonic) force carriers between that matter---we are grouping the latter together with matter in this paper. As hinted at before, the strongest criterion (\hyperref[kc:mass]{\makeAlph{\getrefnumber{kc:mass}}}) thus seems to be overkill; at most criterion \hyperref[kc:tmunu]{\makeAlph{\getrefnumber{kc:tmunu}}} should be sufficient for labeling something matter. Depending on how much one lowers the bar for sufficiency the metric field in GR might also be considered matter according to these criteria. Regardless of where the bar lies exactly, the scalar field added in Berezhiani and Khoury's Superfluid Dark Matter Theory (SFDM), satisfying even the strongest criterion, would definitely count as matter, making it a Dark Matter theory as their choice of name suggests! One issue though is that it may be unclear, especially in the context of our case study, what is meant exactly by having mass. \todo{IGNORE: I've recently gotten a bit confused about this point; I've emailed Berezhiani.}Particle physicists may equate mass to the pole of the propagator. In gravitational physics, the total mass of a gravitating system like a star or a black hole is defined as its Komar, ADM or Bondi mass, which include contributions both from the gravitational binding energy holding the star (say) together, and from the matter fields that make it shine. However, these three conceptions of mass don't necessarily coincide, and can only be defined in special spacetimes.\footnote{Both the ADM mass and the Bondi mass rely on approximating the spacetime containing the star or black hole as asymptotically flat, and rely on the resulting Killing symmetries to define the total mass of the body in question. The Bondi mass is defined via asymptotic symmetries at spatial infinity, the ADM mass via those at null infinity. The Komar mass does not demand asymptotic flatness but stationarity, and so here too  Killing symmetries do much of the work of defining a concept of mass.} Nevertheless, this has not prevented condensed matter physicists, whose discipline inspired core elements of SFDM, from using the concept of mass. 

That being said, it is also true that there are regions within galaxies where the mass-energy-momentum associated with $\Phi$ plays a subdominant or even negligible role in accounting for the (MONDian) behaviour of luminous matter in galaxies. The explanatory story can then be told (almost) fully in terms of effective metrics. Moreover, even Rovelli admits that, although the metric field in GR has adopted most of the classic properties of matter, this ``is not to say that the gravitational field is \emph{exactly} the same object as any other field. The very fact that it \emph{admits} an interpretation in geometrical terms witnesses to its peculiarity'' \cite[p.194]{rovelli1997}. Before passing judgement it therefore seems advisable to turn to the second family of criteria popular in the literature, the spacetime criteria.

\subsection{Spacetime evaluation of SFDM} \label{geoFDM}

%\todo{Consistently use either latin or greek indices.}

%\todo{Hmm, we may need to distinguish beteen $\Phi$ and $\theta$.}
In this subsection we evaluate the SFDM scalar field $\Phi$ according to the spacetime criteria. Crucial to this discussion will be the fact that in SFDM, within the superfluid regime, normal matter (i.e.~luminous, non-dark matter) `feels' the effective metric $\tilde{g}_{\mu\nu}^{SFDM}$---built up out of the Einstein metric and the SFDM scalar field---instead of just the Einstein metric $g_{\mu\nu}$. For, in that regime, $\mathcal{L}_{lum-mat}$ is a function of $\tilde{g}_{\mu\nu}^{SFDM}$ and of covariant derivatives with respect to that effective metric (\hyperref[crucial]{Eq.\ref{crucial}}), rather than a function of $g_{\mu\nu}$ and of covariant derivatives with respect to that Einstein metric\todo{This stuff needs to be explained in the FDM section. Also make sure that I use the same notation for the metrics there, and the same terminology (i.e. effective/physical).}:\footnote{See \hyperref[newFDM]{fn.\ref{newFDM}}.}$^,$\footnote{We assume the simplest possible case, i.e.~minimal coupling and only first order derivatives of the matter fields, which is typical for a matter Lagrangian.}
\begin{equation} \label{crucial2}
\mathcal{L}_{lum-mat} = \mathcal{L}(\tilde{g}^{SFDM}_{\mu\nu},\psi^{\alpha},\psi^{\alpha}_{;\mu | \tilde{g}^{SFDM}}).
\end{equation}

Let us start with the `strong geodesic criterion'. We first turn to test particles, followed by a discussion of the behaviour of light rays. According to SFDM, do massive test particles (of luminous matter)\footnote{We are here not considering test matter made out of $\Phi$, but see the section on breakdown interpretations in the sequel paper \cite{martenslehmkuhl2}.} follow geodesics, and if so, the geodesics of which metric? The Einstein metric? The effective SFDM metric? A different metric altogether? Let us recall the geodesic theorem by Geroch and Jang \cite{geroch1975}. Suppose that given any open subset $O$ of manifold $M$ with metric $g'_{\mu\nu}$ containing  a curve $\gamma$, there exists a smooth, symmetric field $T^{\mu\nu}$ with the following properties:
\begin{enumerate} 
	\item $T^{\mu\nu}$ satisfies the \emph{strengthened dominant energy condition}, i.e. given any timelike vector $\xi^{\mu}$ at any point in $M$, $T^{\mu\nu}\xi_{\mu}\xi_{\nu} \geq 0$, and either $T^{\mu\nu} = 0$ or $T^{\mu\nu}\xi_{\nu}$ is timelike;\footnote{Though the metric does not explicitly appear in the definition of the strengthenend dominant energy condition, the latter nevertheless can only be defined with reference to at least conformal, if not metric, structure. See \cite[\S 5.2]{lehmkuhl2011} for details.}
	\item $T^{\mu\nu}$ satisfies the \emph{conservation condition}, i.e.\ $\nabla_{\nu} T^{\mu\nu}= \mathbf{0}$, where the covariant derivatives are with respect to $g'_{\mu\nu}$;
	\item supp($T^{\mu\nu}$) $\subset O$; and
	\item there is at least one point in $O$ at which $T^{\mu\nu} \neq \mathbf{0}$.
\end{enumerate}
Then $\gamma$ is a timelike curve that may be reparametrized as a geodesic. \qed

The paths of test particles are supposed to correspond to $\gamma$ in virtue of condition 4, whereby the non-vanishing of $T^{\mu\nu}$ is taken as corresponding to the presence of a massive test particle. However, note that there is no further requirement that $T^{\mu\nu}$ be the stress-energy-momentum tensor, though it needs to be a second-rank tensor that can be interpreted as an indicator of the presence of matter in order to use the theorem for making predictions about the presence of matter. The theorem was further generalized by Geroch and Ehlers \cite{geroch2003} and by Geroch and Weatherall  \cite{geroch2018motion}, but for our purposes the original Geroch-Jang theorem suffices.\footnote{See also \cite{weatherall2017} for further discussion of the Geroch-Jang theorem and its conditions and applications, and \cite{Lehmkuhl:2017a} for  arguments why Einstein preferred a different type of geodesic theorem.}

Our question now becomes: for which metric $g'_{\mu\nu}$ within SFDM does the geodesic theorem hold? In particular, with respect to (the covariant derivatives associated with) which metric is $T^{\mu\nu}$ conserved? For that is condition 2 of the Geroch-Jang theorem, and drives much of the proof of the theorem, telling us on the geodesics of which metric test particles move.  To that end, recall the following further result. Consider any matter field, describable by an action that is a function (only) of that field, some metric $g'_{\mu\nu}$ and covariant derivatives of that field with respect to that metric. If that action is required to be diffeomorphism invariant, it can be shown that there exists a rank-2 tensor $T_{\mu\nu}$ associated with that matter field which is symmetric, vanishes in open regions only if the field configuration (satisfying the field equations) vanishes there,\footnote{As long as the Lagrangian density vanishes only where the field (satisfying the field equations) does and has non-trivial dependence on the metric (configuration satisfying the field equations for that metric).} \todo{CAN'T FIND THIS: Check Dennis' comments in 10page document} and which is covariantly conserved with respect to $g'_{\mu\nu}$ (in virtue of the matter field equations holding) \cite[p.64--67]{hawkingellis1973} \cite[p.456]{wald1984}. 

The luminous-matter part of the SFDM action (applicable to the superfluid regime) has exactly the required form, with $g'_{\mu\nu} = \tilde{g}_{\mu\nu}^{SFDM}$ (\hyperref[crucial2]{Eq.\ref{crucial2}}). From the requirement of diffemorphism invariance of this action, one thus obtains a symmetric rank-2 tensor $T_{\mu\nu}$ which, under the additional assumption of it satisfying the strengthened dominant energy condition, satisfies the geodesic theorem with respect to $g_{\mu\nu}^{SFDM}$. Thus, within the superfluid regime, test particles (of luminous matter) in SFDM follow the timelike geodesics of the effective metric $g_{\mu\nu}^{SFDM}$! The SFDM scalar field modifies the behaviour of those test particles, in the sense that they do not follow the timelike geodesics of the Einstein metric, as they do in GR, but instead those of an effective metric built up out of the Einstein metric and that SFDM scalar field.
\todo{CAN'T FIND THIS: Incorporate my worries on p10-11 of the 13page notes, including Dennis' comments. Also, please break up the previous few sentences into shorter ones, very hard to read as they stand.}

To determine the behaviour of light rays in SFDM, the most standard way would be to consider the geometrical-optical limit of electromagnetic fields defined on an SFDM spacetime. As of yet, no-one has worked this out in any detail. However, what drives the argument  that light rays move on null geodescis in the context of GR is that one  assumes minimal coupling between the metric and the electromagnetic field, i.e. replaces the partial derivatives in the Maxwell equations by covariant derivatives, where the covariant derivative is compatible with the metric that solves the gravitaional field equations, and with the help of which the curvature tensor featuring in the field equations is defined. One then derives the  wave equation for the electromagnetic vector potential on a curved background, which features a contraction of said curvature tensor. Then one solves this wave equation approximately by introducing a parameter $\epsilon$ that tracks how rapidly various terms approach zero or infinity in said approximation,\footnote{See e.g. \cite{misnerthornewheeler1973}, p.570-577 for details.} and then observes that looking at terms of order $O(\frac{1}{\epsilon})$ and $O(\frac{1}{\epsilon^2})$ one can derive that in that limit, where electromagnetic waves can be approximated as (light) rays, the light rays move on null geodesics of the connection. Now, if one assumes that in the SFDM context the partial derivatives occuring in the Maxwell equations on flat spacetime should be replaced by covariant derivatives with respect to the effective metric $\tilde{g}_{\mu \nu}$ from Eq.\ref{gFDM} (cf.~Eq.\ref{crucial2}), then it seems plausible that one could make an analogous argument in the context of SFDM.\footnote{However, note that the curvature tensor derived from $\tilde{g}_{\mu \nu}$ would contain the phase of the newly introduced superfluid field, $\theta$, as well as the velocity vector field $u_\mu$ of its normal fluid component. Since part of the assumption that goes into the above approximation scheme is that the `typical reduced wavelength' of the electromagnetic waves is small compared to the `typical component of the Riemann tensor in a typical Lorentz frame', it could be that the more complicated curvature tensor gets in the way of imposing these conditions.}  If this is true, then light rays in SFDM would follow geodesics of the effective metric $\tilde{g}_{\mu \nu}$. 

The effective metric in SFDM (plus Maxwell) would then be \emph{as} spatiotemporal as the Einstein metric in GR (plus Maxwell) as far as the behaviour of light rays is concerned. \todo{IGNORE: Mention that in the later paper with Benoit they choose gamma such that the metric becomes conformal to GR, in which the null geodesics of the Einstein metric and effective metric coincide! See email with Berezhiani for their motivations for doing so. In their later paper they'll probalby also mention the neutron merger as a motivation(check this!). Edit: on p3 of the 2017 paper they seem to suggest that photons don't just couple to a metric conformal to the Einstein metric, but simply to the Einstein metric itself. Note that they seem to decouple this choice from the choice of which metric couples to baryonic luminous matter! In order to get the MOND result, they probably still need to couple baronic matter to the/a effective metric, although they may of course still set gamma such that that metric is conformal, even though such a choice can't be motivated anymore by gravitational lensing. Edit: see also page with handwritten notes on important points of berezhiani 2016 and 2017, written at the end of October 2018.}%\todo[color=white]{\niels{Dennis: Do we urgently need this paragraph on null geodesics? None of us has studied the geometric-optical limit in GR, and the argument that it carries over to SFDM via the mere form of the EM Lagrangian seems far too quick to me.}}

%Again, since the form of the matter Lagrangian (in this case the Lagrangian governing the Faraday tensor) is the same, other than being with respect to the physical metric of that theory rather than the Einstein metric, the matter equations (i.e./ the Maxwell equations) will be the same (but then with reference to the physical metric). The difference is of course that the space of physical metric solutions differs (in general) from the space of (Einstein metric) solutions in standard GR. And even in GR there is no general proof that in the geometrical-optical limit the light rays will follow null geodesics---G\"{o}del spacetimes being an exception \cite{menonetal2018}. Similarly, whether light rays follow null geodesics of the physical metric would have to be determined for each solution separately. 

Finally, we move to the chronogeometricity criterion.\footnote{Berezhiani, Khoury and Famaey \cite{berezhiani2017} assert that MOND, and by extension SFDM, violates the `strong equivalence principle'. It is clear though that what they refer to as (an implication of) the `strong equivalence principle', namely that ``a homogeneous acceleration has no physical consequence'' \cite[p.14]{berezhiani2017} (which would indeed be contradicted by MOND's acceleration scale $a_0$), is not what we have defined as the strong equivalence principle. We would call this a version of the Einstein equivalence principle \cite{lehmkuhlequivalenceprinciples}.} \todo{Dennis, does the EEP play any role in our paper? I'd say not.... right?} To determine the local validity of Special Relativity, we need to determine whether any curvature terms pop up in the equations of motions of the luminous matter (that makes up rods and clocks). To this end, compare the way luminous matter is coupled universally to the Einstein metric $\g$ within General relativity, 
\begin{equation}
\mathcal{L}_{lum-mat-in-GR} = \mathcal{L}_{lum-mat}(\g,\psi^{\alpha},\psi^{\alpha}_{;\mu | \g})
\end{equation}
with the way in which the direct analogue of that Lagrangian term would be included in (the superfluid regime of) SFDM, that is by a universal coupling to the SFDM metric, \hyperref[crucial2]{Eq.\ref{crucial2}}. As the matter equations are determined by varying $S_{lum-mat}$, which has the same form for both theories except for the role of the metric being played by different tensors, these equations of motions also have the same form except for any potential curvature terms being with respect do different metric tensors (i.e.~$\g$ and $\tilde{g}_{\mu\nu}^{SFDM}$ respectively). Strict satisfaction of the chronogeometricity criterion would require there to be no curvature terms at all in the local equations of motions of the luminous matter fields $\psi^{\alpha}$. It turns out however that that not only non-minimal coupling can generate curvature terms \cite[\S 9.4.1]{brown2005}, but that even minimal coupling can do so \cite{readbrown2018,wald1984}, thereby strictly violating the chronogeometricity criterion. The approximate validity of the criterion then depends on whether one can effectively ignore the curvature terms in the specific experimental or operational context one is interested in. In a general sense, the effective metric in the context of SFDM is thus not worse off than the Einstein metric in the context of GR when it comes to satisfying the chronogeometricity criterion. Within each theory one would have to check, for each solution and for (each neighbourhood of) each event in that solution, whether the curvature term (with respect to the respective metric) is sufficiently small to provide Special Relativity up to the required approximation.

In conclusion, as far as the (strong) geodesic and chronogeometricity criteria are concerned, the effective SFDM metric, constructed out of the Einstein metric and the new SFDM scalar field, satisfies them, within the superfluid regime, to the same extent that the Einstein metric does within GR. If the Einstein metric within GR deserves to be labeled `spacetime', so would the SFDM scalar deserve to be labeled an `aspect' or `modification' of spacetime.

%In a Lagrangian theory, the equations of motion of the matter fields are obtained from the action by varying with respect to those fields. If the matter Lagrangian part is exactly the same as in GR, with the replacement of $g$ by $\tilde{g}$ and the corresponding covariant derivatives (\hyperref[grform]{Eq.\ \ref{grform}}), as it is in all of the above theories, then we will get exactly the same equations of motion, but now with respect to the new metric. The SEP will then hold to the same extent as in GR+matter theories. Even just in that GR+matter context, we know that not only non-minimal coupling can generate curvature terms \cite{brown2005}, but that even minimal coupling can do so \cite{readbrown2018,wald1984}, thereby strictly violating the SEP. The approximate validity of the SEP then depends on whether you can effectively ignore the curvature terms in the specific experimental context you are in. The same will apply to the other theories discussed above, although the curvature terms will now be referring to the curvature of $\tilde{g}$. One difference, though, is that the space of solutions of $\tilde{g}$ of those theories will of course differ from the space of metric solutions to GR.

\section{Conclusion \& Outlook}

Let us sum up where that leaves us so far. The scalar field $\Phi$ introduced by SFDM satisfies the matter criterion of (almost) the highest strength. Regardless of where one draws the line, that is which matter strength one considers sufficient for something being matter, the scalar field $\Phi$ will clearly count as a matter field. We have also seen that, at least within the superfluid regime, the SFDM effective metric, constructed out of the Einstein metric and $\Phi$, satisfies the spacetime criteria to the same extent as the Einstein metric does within GR. $\Phi$ is thus---at least for temperatures below the critical temperature for superfluidity---also as much of an aspect of spacetime as one can expect of a dynamical field. 

The significance of this result becomes clear when we contrast it with a previous attempt by Khoury to account for dark phenomena \cite{khoury2015}. That theory adds not one but two scalar fields. According to Khoury, the first scalar field ``behaves as a dark matter fluid on large scales''\cite[p.1]{khoury2015}. The second ``mediates a fifth force that modifies gravity on nonlinear scales''\cite[p.1]{khoury2015}. Even if these quotes hold up under the interpretations offered by our families of criteria, it would be one field being responsible for satisfying one or more of the matter criteria, and the other field being responsible for ticking off the spacetime criteria. This is neither novel nor especially interesting with respect to the interpretational questions posed in our two companion papers: both within (field-theoretic versions of) Newtonian physics and Special Relativity do the metric fields satisfy the spacetime criteria (and none of the matter criteria) and, say, electric charge density fields satisfy the matter criteria (and none of the spacetime criteria). What is interesting about SFDM is that, rather than requiring a second field, the modification of the metric is associated with a four-velocity of the only new (scalar) field $\Phi$ which also plays the dark matter role; the ``DM and MOND [or Modified Gravity] components have a common origin, representing different phases of a single underlying substance''\cite[p.1,3]{berezhiani2016}.%\todo{We might want to look into other hybrid theories cited in \cite{berezhiani2016}.} 

The result of this paper still leaves open the interpretational questions mentioned at the beginning. What follows from this for the distinction between dark matter and modified gravity, as well as the broader distinction between matter and spacetime, both within SFDM and in general? Does the fact that $\Phi$ seems to satisfy the spacetime criteria only below the critical temperature for superfluidity imply that it is `more' of a dark matter field than it is a modification of spacetime, or are both roles on a par? These and other remaining interpretational questions will be the focus of the companion paper \cite{martenslehmkuhl2}.

\section*{Acknowledgements}

We would like to acknowledge support from the DFG Research Unit ``The Epistemology of the Large Hadron Collider'' (grant FOR 2063). Within this research unit we are particularly indebted to the other members of our `LHC, Dark Matter \& Modified Gravity' project team---Miguel \'{A}ngel Carretero Sahuquillo, Michael Kr\"{a}mer and Erhard Scholz---for invaluable and extensive discussions and comments on many iterations of this paper. We would furthermore like to thank Radin Dardashti, Tushar Menon, James Read, Joshua Rosaler, Kian Salimkhani, Michael St\"{o}ltzner and Adrian W\"{u}thrich for valuable discussions and comments, as well as the audiences of the Geneva Symmetry Group Research Seminar (Geneva, Switzerland, 2018), the Fifth International Conference on the Nature and Ontology of Spacetime (Albena, Bulgaria, 2018), the British Society for the Philosophy of Science Conference (Oxford, UK, 2018), the Philosophy of Science \& Technology Colloquium (Aachen, Germany, 2018), the Philosophy Colloquium (Cologne, Germany, 2018), the Dark Matter \& Modified Gravity Conference (Aachen, Germany, 2019), the German Society for the Philosophy of Science Conference (Cologne, Germany, 2019) and the First Oxford-Notre Dame-Bonn Workshop on the Foundations of Spacetime Theories (Oxford, UK, 2019).

\bibliographystyle{unsrt}
\bibliography{bibdmmg}

\end{document}